\documentclass[12pt]{article}
\usepackage{graphicx}
\usepackage{lscape}
\usepackage{citesort}
\usepackage{amssymb}
\usepackage{appendix}
\usepackage{multirow}

\begin{document}
\def\qq{\langle \bar q q \rangle}
\def\uu{\langle \bar u u \rangle}
\def\dd{\langle \bar d d \rangle}
\def\sp{\langle \bar s s \rangle}
\def\GG{\langle g_s^2 G^2 \rangle}
\def\Tr{\mbox{Tr}}
\def\figt#1#2#3{
        \begin{figure}
        $\left. \right.$
        \vspace*{-2cm}
        \begin{center}
        \includegraphics[width=10cm]{#1}
        \end{center}
        \vspace*{-0.2cm}
        \caption{#3}
        \label{#2}
        \end{figure}
    }

\def\figb#1#2#3{
        \begin{figure}
        $\left. \right.$
        \vspace*{-1cm}
        \begin{center}
        \includegraphics[width=10cm]{#1}
        \end{center}
        \vspace*{-0.2cm}
        \caption{#3}
        \label{#2}
        \end{figure}
                }

\def\ds{\displaystyle}
\def\beq{\begin{equation}}
\def\eeq{\end{equation}}
\def\bea{\begin{eqnarray}}
\def\eea{\end{eqnarray}}
\def\beeq{\begin{eqnarray}}
\def\eeeq{\end{eqnarray}}
\def\ve{\vert}
\def\vel{\left|}
\def\ver{\right|}
\def\nnb{\nonumber}
\def\ga{\left(}
\def\dr{\right)}
\def\aga{\left\{}
\def\adr{\right\}}
\def\lla{\left<}
\def\rra{\right>}
\def\rar{\rightarrow}
\def\lrar{\leftrightarrow}
\def\nnb{\nonumber}
\def\la{\langle}
\def\ra{\rangle}
\def\ba{\begin{array}}
\def\ea{\end{array}}
\def\tr{\mbox{Tr}}
\def\ssp{{\Sigma^{*+}}}
\def\sso{{\Sigma^{*0}}}
\def\ssm{{\Sigma^{*-}}}
\def\xis0{{\Xi^{*0}}}
\def\xism{{\Xi^{*-}}}
\def\qs{\la \bar s s \ra}
\def\qu{\la \bar u u \ra}
\def\qd{\la \bar d d \ra}
\def\qq{\la \bar q q \ra}
\def\gGgG{\la g^2 G^2 \ra}
\def\q{\gamma_5 \not\!q}
\def\x{\gamma_5 \not\!x}
\def\g5{\gamma_5}
\def\sb{S_Q^{cf}}
\def\sd{S_d^{be}}
\def\su{S_u^{ad}}
\def\sbp{{S}_Q^{'cf}}
\def\sdp{{S}_d^{'be}}
\def\sup{{S}_u^{'ad}}
\def\ssp{{S}_s^{'??}}

\def\sig{\sigma_{\mu \nu} \gamma_5 p^\mu q^\nu}
\def\fo{f_0(\frac{s_0}{M^2})}
\def\ffi{f_1(\frac{s_0}{M^2})}
\def\fii{f_2(\frac{s_0}{M^2})}
\def\O{{\cal O}}
\def\sl{{\Sigma^0 \Lambda}}
\def\es{\!\!\! &=& \!\!\!}
\def\ap{\!\!\! &\approx& \!\!\!}
\def\md{\!\!\!\! &\mid& \!\!\!\!}
\def\ar{&+& \!\!\!}
\def\ek{&-& \!\!\!}
\def\kek{\!\!\!&-& \!\!\!}
\def\cp{&\times& \!\!\!}
\def\se{\!\!\! &\simeq& \!\!\!}
\def\eqv{&\equiv& \!\!\!}
\def\kpm{&\pm& \!\!\!}
\def\kmp{&\mp& \!\!\!}
\def\mcdot{\!\cdot\!}
\def\erar{&\rightarrow&}
\def\olra{\stackrel{\leftrightarrow}}
\def\ola{\stackrel{\leftarrow}}
\def\ora{\stackrel{\rightarrow}}

\def\simlt{\stackrel{<}{{}_\sim}}
\def\simgt{\stackrel{>}{{}_\sim}}


\title{
         {\Large
                 {\bf
                      Mass and Meson-Current Coupling Constant of the Tensor $D_2^*(2460)$
                 }
         }
      }

\author{\vspace{1cm}\\
{\small H.
Sundu$^1$ \thanks
{e-mail: hayriye.sundu@kocaeli.edu.tr}\,\,, K. Azizi$^2$ \thanks {e-mail: kazizi@dogus.edu.tr}}  \\
{\small $^1$ Department of Physics , Kocaeli University, 41380
Izmit, Turkey}\\
{\small $^2$ Department of Physics, Do\u gu\c s University,
Ac{\i}badem-Kad{\i}k\"oy, 34722 Istanbul, Turkey} }
\date{}

\begin{titlepage}
\maketitle
\thispagestyle{empty}

\begin{abstract}
We calculate the mass and meson-current coupling constant of the $D_2^*(2460)$ tensor
meson in the framework of QCD sum rules. The obtained result on the
mass is compatible with the experimental data.
\end{abstract}

~~~PACS number(s): 11.55.Hx, 14.40.Lb
\end{titlepage}

\section{Introduction}

The  semileptonic $B$ meson transitions to the orbitally excited charmed mesons as well as the strong transitions of the excited charmed mesons into the other charmed states have been
in focus of attention of many Collaborations in the recent years. The  BaBar and Belle Collaborations  reported
the measurement of the  products/ratios of the branching fractions of some  semileptonic and hadronic decays of the $B$ meson and orbitally excited charmed meson channels \cite{Babar1,Babar2,Belle1}.
Considering these experimental progress and the fact that the decay channels containing orbitally excited charmed meson in the final state supply a considerable contribution to the total semileptonic
$B$ meson decay width, more knowledge about the properties of orbitally excited charmed mesons like $D_2^*(2460)$ is needed both experimentally and theoretically.

In the present letter, we calculate the mass and  meson-current coupling constant of the $D_2^*(2460)$ tensor meson with quantum numbers $I(J^{P})=1/2~(2^{+})$ in the framework of  the  QCD sum rules
as one of the most applicable
 and powerful non-perturbative approaches to hadron physics. For details about the method and some of its  applications see for instance  \cite{MAShifman1,MAShifman2,L.J.Reinders,colangelo}.
Our results on the mass and meson-current coupling constant of  $D_2^*(2460)$ tensor meson can be used
in theoretical calculations on the decays of heavy mesons  into orbitally excited charmed $D_2^*(2460)$ meson. Moreover, they may be used to analyze the electromagnetic, semileptonic and strong  decay modes
 of the $D_2^*(2460)$ into other charmed
 and  lighter mesons (For some  decay modes of $D_2^*(2460)$ tensor meson and decay modes of heavy mesons into this state see \cite{K.Nakamura}).
Some properties such as mass, meson-current coupling constant and electromagnetic multi-poles of the heavy-heavy, light-heavy and light-light tensor mesons have previously calculated using different frameworks.
For some of them see
\cite{aliev,Kazim1,Kazim2,Chen,Kazim3,Wilcox} and references therein.
Some semileptonic decays of the $B$ meson into the orbitally excited charmed mesons have also been studied in \cite{Segovia} within the framework of  constituent quark
model.

 The letter  is organized as follows. In section II, we briefly present    calculations of the mass and meson-current coupling constant of the  $D_2^*(2460)$ tensor meson within the framework of
the QCD sum rules. Section III is devoted to the
numerical analysis of the considered observables as well as comparison of our result on mass with experimental data.

\section{QCD sum rules for  mass and meson-current coupling constant  of the $D_2^*(2460)$ tensor meson}
This section is dedicated to  calculation of the mass
and meson-current coupling constant of the  $D_2^*(2460)$ tensor meson in the framework of  the  QCD
sum rules. The starting point is to consider  the following two-point correlation function:
\begin{eqnarray}\label{correl.func.101}
\Pi _{\mu\nu,\alpha\beta}=i\int
d^{4}xe^{iq(x-y)}{\langle}0\mid {\cal T}[j _{\mu\nu}(x)
\bar j_{\alpha\beta}(y)]\mid  0{\rangle},
\end{eqnarray}
where, $j_{\mu\nu}$ is the interpolating current of the
$D_2^*(2460)$ tensor meson and ${\cal T}$ is the time ordering
operator. The   current $j_{\mu\nu}$ is written in terms of the quark fields as
\begin{eqnarray}\label{tensorcurrent}
j _{\mu\nu}(x)=\frac{i}{2}\left[\bar u(x) \gamma_{\mu} \olra{\cal
D}_{\nu}(x) c(x)+\bar u(x) \gamma_{\nu}  \olra{\cal D}_{\mu}(x)
c(x)\right],
\end{eqnarray}
where the $ \olra{\cal D}_{\mu}(x)$ denotes the four-derivative with
respect to x acting on the left and right, simultaneously. It is
given as
\begin{eqnarray}\label{derivative}
\olra{\cal D}_{\mu}(x)=\frac{1}{2}\left[\ora{\cal D}_{\mu}(x)-
\ola{\cal D}_{\mu}(x)\right],
\end{eqnarray}
with,
\begin{eqnarray}\label{derivative2}
\overrightarrow{{\cal D}}_{\mu}(x)=\overrightarrow{\partial}_{\mu}(x)-i
\frac{g}{2}\lambda^aA^a_\mu(x),\nonumber\\
\overleftarrow{{\cal D}}_{\mu}(x)=\overleftarrow{\partial}_{\mu}(x)+
i\frac{g}{2}\lambda^aA^a_\mu(x).
\end{eqnarray}
Here,  $\lambda^a$ are the Gell-Mann matrices and $A^a_\mu(x)$ is
the external  gluon fields. Considering the Fock-Schwinger gauge ($x^\mu A^a_\mu(x)=0$), these fields are expressed
in terms of the gluon field strength tensor
\begin{eqnarray}\label{gluonfield}
A^{a}_{\mu}(x)=\int_{0}^{1}d\alpha \alpha x_{\beta} G_{\beta\mu}^{a}(\alpha x)=
\frac{1}{2}x_{\beta} G_{\beta\mu}^{a}(0)+\frac{1}{3}x_\eta x_\beta {\cal D}_\eta
G_{\beta\mu}^{a}(0)+...
\end{eqnarray}
The currents contain derivatives with respect to the
space-time, hence we consider the two currents at points x and y.
After applying the derivatives with respect to  $y$,  we will put $y=0$.

According to the general criteria of the QCD sum rules, the aforementioned correlation function is calculated via two alternative ways: phenomenologically (physical side) and theoretically (QCD side). In physical side,
the two-point correlation function is calculated in terms of  hadronic degrees of freedom like mass, meson-current coupling constant, etc. The QCD side is obtained in terms of the QCD parameters such as quark masses, quark and gluon
 condensates, etc. The two-point QCD sum rules is obtained matching coefficients of the same structure representing the tensor mesons from both sides through a dispersion relation and quark-hadron duality assumption.
Finally, we apply Borel transformation to stamp down the contributions belong to the higher states and continuum.

\subsection{The physical side}

In the  physical side, the correlation function
is obtained inserting complete set of hadronic state having the
same quantum numbers as the interpolating current
$j_{\mu\nu}$  into Eq. (\ref{correl.func.101}).
 After performing  integral over four-x and putting $y=0$, we
obtain the physical side of correlation function as following form:
\begin{eqnarray}\label{phen1}
\Pi _{\mu\nu,\alpha\beta}=\frac{{\langle}0\mid  j _{\mu\nu}(0) \mid
D_2^*(2460)\rangle \langle D_2^*(2460)\mid
\bar j_{\alpha\beta}(0)\mid
 0\rangle}{m_{D_2^*(2460)}^2-q^2}
&+& \cdots,
\end{eqnarray}
 where $\cdots$ represents  contribution of  the higher states and continuum.
To proceed,  we need  to
know the matrix element $\langle 0 \mid j_{\mu\nu}(0)\mid
D_2^*(2460)\rangle$, which is defined in terms of the
meson-current coupling constant, mass and polarization tensor
\begin{eqnarray}\label{lep}
\langle 0 \mid j_{\mu\nu}(0)\mid
D_2^*(2460)\rangle=f_{D_2^*(2460)}
m_{D_2^*(2460)}^3\varepsilon_{\mu\nu}.
\end{eqnarray}
Combining Eq. (\ref{phen1}) and Eq. (\ref{lep}) and performing
summation over polarization tensor via
\begin{eqnarray}\label{polarizationt1}
\varepsilon_{\mu\nu}\varepsilon_{\alpha\beta}^*=\frac{1}{2}T_{\mu\alpha}T_{\nu\beta}+
\frac{1}{2}T_{\mu\beta}T_{\nu\alpha}
-\frac{1}{3}T_{\mu\nu}T_{\alpha\beta},
\end{eqnarray}
with,
\begin{eqnarray}\label{polarizationt2}
T_{\mu\nu}=-g_{\mu\nu}+\frac{q_\mu q_\nu}{m_{D_2^*(2460)}^2},
\end{eqnarray}
the final representation of physical side is obtained as
\begin{eqnarray}\label{phen2}
\Pi
_{\mu\nu,\alpha\beta}=\frac{f^2_{D_2^*(2460)}m_{D_2^*(2460)}^6}
{m_{D_2^*(2460)}^2-q^2}
\left\{\frac{1}{2}(g_{\mu\alpha}~g_{\nu\beta}+g_{\mu\beta}~g_{\nu\alpha})\right\}+
\mbox{other structures}+...,
\end{eqnarray}
where, the explicitly written  structure  gives  contribution to the
tensor state.

\subsection{The QCD side}

The correlation function in QCD  side,  is calculated
in deep Euclidean region, $q^2\ll0$, by the help of  operator product expansion
(OPE) where the short and long distance contributions are separated. The short distance effects
 are calculated using the perturbation theory, while the
long distance effects are parameterized in terms of quark and gluon condensates.

Any coefficient of the structure, $\frac{1}{2}(g_{\mu\alpha}~g_{\nu\beta}+g_{\mu\beta}~
g_{\nu\alpha})$, in QCD side, i.e. $\Pi(q^2)$,  can be written as a dispersion relation
\begin{eqnarray}\label{QCDside}
\Pi (q^2)=\int^{}_{}ds
\frac{\rho(s)}{s-q^2},
\end{eqnarray}
where the spectral density is given by the imaginary part of the
$\Pi(q^2)$ function, i.e.,  $\rho(s)=\frac{1}{\pi}Im[\Pi(s)]$. As we mentioned above, the correlation function contains both perturbative and non-perturbative effects, hence
 the spectral density can be decomposed as
\begin{eqnarray}\label{rho}
\rho(s)=\rho^{pert}(s)+\rho^{nonpert}(s),
\end{eqnarray}
where, $\rho^{pert}(s)$ and $\rho^{nonpert}(s)$ denote the
contributions  coming from perturbative  and
non-perturbative effects, respectively.

Now, we proceed to calculate the spectral density $\rho(s)$. Making use of the  tensor current presented in
 Eq. (\ref{tensorcurrent}) into the correlation function  in
Eq. (\ref{correl.func.101})  and contracting out all quark fields applying the
Wick's theorem, we get:
\begin{eqnarray}\label{correl.func.2}
\Pi _{\mu\nu,\alpha\beta}&=&\frac{i}{4}\int d^{4}xe^{iq(x-y)}
\Bigg\{Tr\left[S_u(y-x)\gamma_\mu\olra{\cal D}_{\nu}(x)
\olra{\cal D}_{\beta}(y)S_c(x-y)\gamma_\alpha\right]\nonumber \\
&&+\left[\beta\leftrightarrow\alpha\right]
+\left[\nu\leftrightarrow\mu\right]+\left[\beta\leftrightarrow\alpha,
\nu\leftrightarrow\mu\right]\Bigg\}.
\end{eqnarray}
To obtain the correlation function from QCD side, we need to know
the heavy  and light quarks propagators $S_c(x-y)$   and
$S_u(x-y)$. These propagators have been calculated in
\cite{L.J.Reinders}. Ignoring the gluon fields which have very small
contributions in our calculations (see also \cite{aliev,Kazim1}),
their explicit expressions between two points  up to quark
condensates can be written as
\begin{eqnarray}\label{heavypropagator}
S_{c}^{ij}(x-y)=\frac{i}{(2\pi)^4}\int d^4k e^{-ik \cdot (x-y)}
\left\{ \frac{\!\not\!{k}+m_c}{k^2-m_c^2}\delta_{ij}
 +\cdots\right\} \, ,
\end{eqnarray}
and
\begin{eqnarray}
S_{u}^{ij}(x-y)&=& i\frac{\!\not\!{x}-\!\not\!{y}}{
2\pi^2(x-y)^4}\delta_{ij}
-\frac{m_u}{4\pi^2(x-y)^2}\delta_{ij}-\frac{\langle
\bar{u}u\rangle}{12}\Big[1 -i\frac{m_u}{4}
(\!\not\!{x}-\!\not\!{y})\Big]\delta_{ij}
\nonumber\\
&-&\frac{(x-y)^2}{192}m_0^2\langle
\bar{u}u\rangle\Big[1-i\frac{m_u}{6}
(\!\not\!{x}-\!\not\!{y})\Big]\delta_{ij} +\cdots \, .
\end{eqnarray}
Note that  the gluon condensates are also ignored in \cite{ZGWang,ZGWang1,ZGWang2}  since their
contributions are suppressed by large denominators, so they
play  minor roles in calculations.

The next step is to put the expressions of the  propagators and apply
the derivatives with respect to x and y in  Eq.
(\ref{correl.func.2}) and finally set  $y=0$.  As a result, the
following  expression for the QCD side of the correlation
function in coordinate space is obtained:
\begin{eqnarray}\label{correl.func.3}
\Pi _{\mu\nu,\alpha\beta}&=&\frac{N_c}{16}
 \int\frac{d^4k}{(2\pi)^4}\frac{1}{k^2-m_c^2}\int
d^{4}xe^{i(q-k)\cdot
x}\Big\{\Big[Tr\Gamma_{\mu\nu,\alpha\beta}\Big]+
\Big[\beta\leftrightarrow\alpha\Big]+\Big[\nu\leftrightarrow\mu\Big]
\nonumber\\&+&
\Big[\beta\leftrightarrow\alpha,\nu\leftrightarrow\mu\Big]\Big\},
\end{eqnarray}
where $N_c=3$ is the color factor and ,
\begin{eqnarray}\label{fonk}
\Gamma_{\mu\nu,\alpha\beta}&=&
k_{\nu}k_{\beta}\Big[\frac{i\!\not\!{x}}{2\pi^2 x^4}
+\Big(\frac{1}{12}+\frac{x^2}{192}m_0^2\Big)\langle
\bar{u}u\rangle\Big]\gamma_{\mu}(\!\not\!{k}+m_c)\gamma_{\alpha}
\nonumber\\
&-&ik_{\nu} \Big[\frac{i}{2
\pi^2}\Big(\frac{\gamma_{\beta}}{x^4}-\frac{4x_{\beta}
\!\not\!{x}}{x^6}\Big)+\frac{x_{\beta}}{96}m_0^2\langle
\bar{u}u\rangle\Big]\gamma_{\mu}(\!\not\!{k}+m_c)\gamma_{\alpha}
\nonumber\\
&+&ik_{\beta}\Big[\frac{i}{2
\pi^2}\Big(\frac{4x_{\nu}\!\not\!{x}}{x^6}
-\frac{\gamma_{\nu}}{x^4}\Big)+\frac{x_{\nu}}{96}m_0^2\langle
\bar{u}u\rangle\Big]\gamma_{\mu}(\!\not\!{k}+m_c)\gamma_{\alpha}
\nonumber\\
&+&\Big[\frac{8i}{2\pi^2}\Big(\frac{ x_{\beta}
x_{\nu}\!\not\!{x}}{x^{8}}+\frac{1}{2x^{6}}
\Big(\delta^{\nu}_{\beta}\!\not\!{x}-\gamma_{\nu}
x_{\beta}+\gamma_{\beta} x_{\nu}\Big) +\frac{ \gamma_{\nu}
x_{\beta}}{x^6}-\frac{4 x_{\beta}x_{\nu}\!\not\!{x}}{x^8}\Big)
\nonumber\\
&+& \frac{\delta^{\nu}_{\beta}}{96}m_0^2\langle
\bar{u}u\rangle\Big]\gamma_{\mu} (\!\not\!{k}+m_c)\gamma_{\alpha} +
\left[\beta\leftrightarrow\alpha\right]+\left[\nu\leftrightarrow\mu\right]+
\left[\beta\leftrightarrow\alpha,\nu\leftrightarrow\mu\right].
\end{eqnarray}
In order to
calculate the integrals, first we transform the terms containing $\frac{1}{(x^2)^n}$    to the momentum space ($x\rightarrow p$)
 and replace $x_{\mu}\rightarrow -i\frac{\partial}{\partial q_{\mu}}$. The integral over four-x gives us a Dirac Delta function which help us perform
the integral over four-k. The last integral which is over four-p is performed  using the
Feynman parametrization and the relation
\begin{eqnarray}\label{Int}
\int d^4p\frac{(p^2)^{\beta}}{(p^2+L)^{\alpha}}=\frac{i \pi^2
(-1)^{\beta-\alpha}\Gamma(\beta+2)\Gamma(\alpha-\beta-2)}{\Gamma(2)
\Gamma(\alpha)[-L]^{\alpha-\beta-2}}.
\end{eqnarray}
After dimensional regularization and taking the imaginary part and selecting the coefficient of the aforesaid structure,  the spectral densities   are obtained as:
\begin{eqnarray}\label{Rhopert}
\rho^{pert}(s)&=&
 \frac{N_c}{960~\pi^2s^3}\Big(m_c^2-s\Big)^4 \Big(2m_c^2+3s\Big),
\end{eqnarray}

and
\begin{eqnarray}\label{rho}
\rho^{nonpert}(s)=-\frac{N_c}{48 s}m_c m_0^2\langle \bar{u}u\rangle.
\end{eqnarray}

After achieving the correlation function in two different ways,  we
match these two different representations  to obtain two-point QCD sum rules for
the  meson-current coupling constant and mass. In order to suppress contributions of
the higher states and continuum, we apply Borel transformation with
respect to the initial momentum squared,  $q^2$, to both sides of
the sum rules and use the quark-hadron duality assumption. As a
result, the following sum rule for the meson-current coupling constant of the
$D_2^*(2460)$ tensor meson is obtained:
\begin{eqnarray}\label{rhomatching}
f^2_{D_2^*(2460)} e^{-m_{D_2^*(2460)}^2/M^2}
=\frac{1}{m_{D_2^*(2460)}^6}\int_{m_c^2}^{s_0} ds
\Big(\rho^{pert}(s)+\rho^{nonpert}(s)\Big)e^{-s/M^2},
\end{eqnarray}
where $s_0$ is the continuum threshold and $M^2$ is the Borel mass parameter. Differentiating Eq.
(\ref{rhomatching}) with respect to $-\frac{1}{M^2}$ and dividing  both sides of the obtained result to both sides of Eq. (\ref{rhomatching}), we  obtain the following
sum rule for the mass of the $D_2^*(2460)$ tensor meson:
\begin{eqnarray}\label{rhomass}
m_{D_2^*(2460)}^2
=\frac{\int_{m_c^2}^{s_0}ds\Big(\rho^{pert}(s)+\rho^{nonpert}(s)\Big)~s~e^{-s/M^2}}
{\int_{m_c^2}^{s_0}ds\Big(\rho^{pert}(s)+\rho^{nonpert}(s)\Big)
e^{-s/M^2}}.
\end{eqnarray}

\section{Numerical results}
In this section, we carry out  numerical analysis of the sum
rules for the mass and meson-current coupling constant of the $D_2^*(2460)$ tensor
meson.  The input parameters are taken to be
$m_c=(1.27^{+0.07}_{-0.09})~GeV$ \cite{K.Nakamura}, $\langle \bar
uu(1~GeV)\rangle=-(0.24\pm0.01)^3~GeV^3$ \cite{B.L.Ioffe} and
$m_0^2(1~GeV) = (0.8\pm0.2)~GeV^2$ \cite{m02}. The
 sum rules contain also two auxiliary parameters, namely the Borel
 parameter $M^2$
and continuum threshold  $s_0$. According to the standard procedure
in QCD sum rules, the physical quantities should be independent of
these parameters, hence we shall look for working regions for these
parameters such that the mass and meson-current coupling constant of the
$D_2^*(2460)$ tensor meson  weakly depend on these helping
parameters. The continuum threshold $s_{0}$ is not totally arbitrary
and it is in correlation with the energy of the first exited state. Since we have no experimental information on the first excited state of the meson under consideration,
comparing to the other charmed mesons with similar mass
we choose the interval $s_0=(8.1\pm0.5)~GeV^2$ for the continuum
threshold (for more information see \cite{dsj,dscol}). This interval is obtained from
$(m_{meson}+0.3)^2\leq s_0\leq(m_{meson}+0.5)^2$ and our numerical calculations show that in this region, the results of physical observables very weakly depend on the continuum threshold.

The working region for the Borel mass parameter is
determined requiring that not only  the higher state and continuum
contributions are suppressed but also the contributions of the higher
order operators are ignorable, i.e. the sum rules are convergent. As
a result, the working region for the Borel parameter is found to be
$ 3~ GeV^2 \leq M^2 \leq 6~ GeV^2 $. To show how the OPE is convergent and the contributions of the higher states and continuum are suppressed in the used Borel window, as an example we plot the result of
sum rule for
the meson-current coupling constant in terms of Borel mass parameter at $s_0=8.2~ GeV^2$ in figure 1. From this figure we see that the main contribution comes from the perturbative part such that the
non-perturbative
part constitutes only $15\%$ of the total contribution. From this figure it is also clear that not only the contribution of the higher states and continuum is zero but also the higher dimensional operators ($d>5$)
have very small contributions and can be safely ignored.
\begin{figure}[h!]
\begin{center}
\includegraphics[width=8cm]{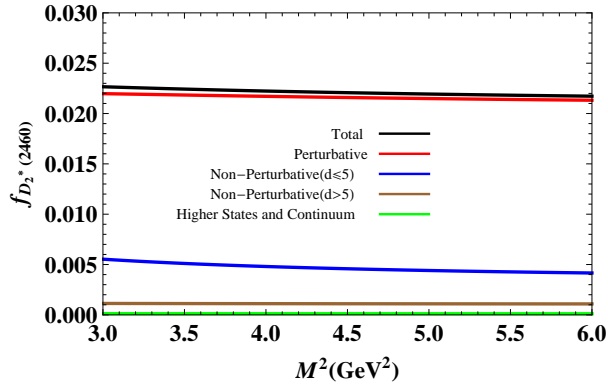}
\end{center}
\caption{Comparison of the different contributions to the sum rule for the meson-current coupling constant at $s_0=8.2~ GeV^2$.}
\label{Diagrams1}
\end{figure}

The dependencies of the mass and total meson-current coupling constant of the $D_2^*(2460)$ tensor meson on Borel mass parameter are also presented in figure 2. From this figure, we
 see  good stability of the observables under consideration with respect to the variation of Borel mass parameter in its working region.
\begin{figure}[h!]
\begin{center}
\includegraphics[width=8cm]{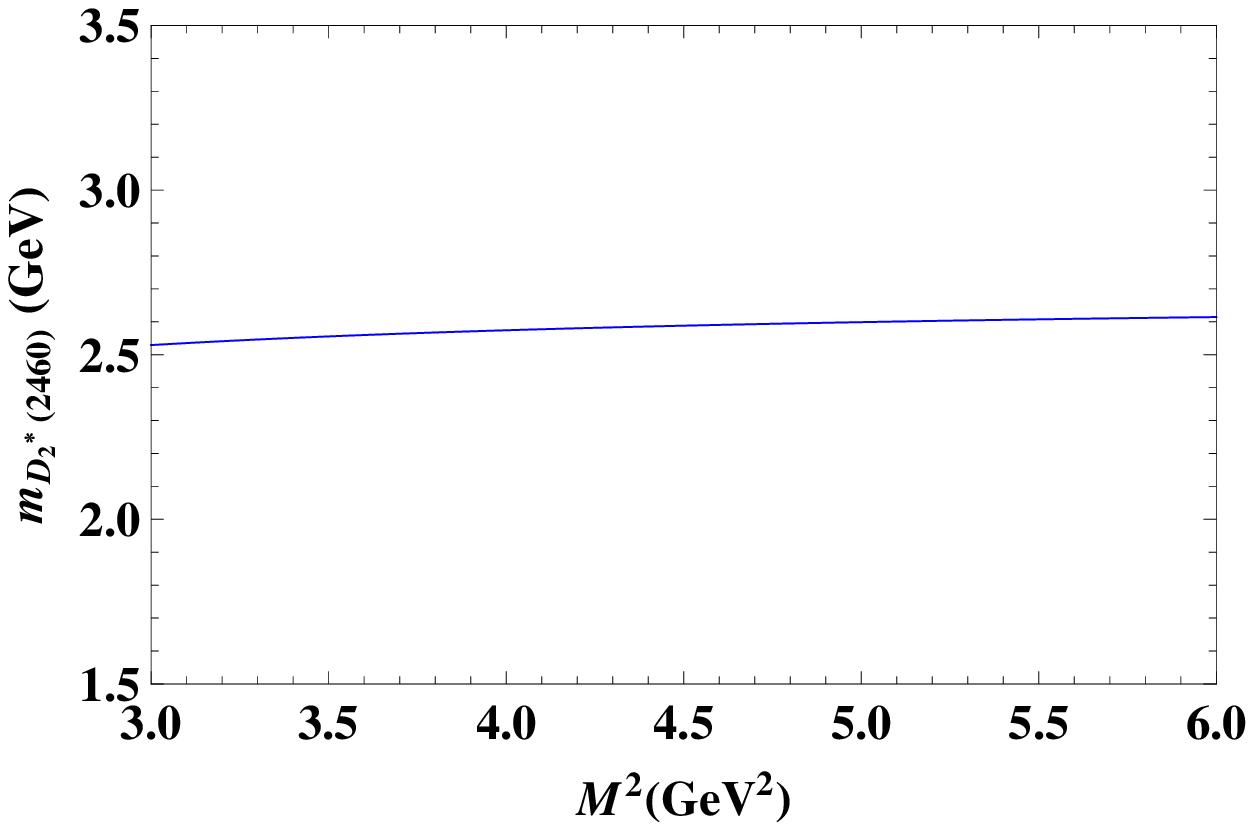}
\includegraphics[width=8cm]{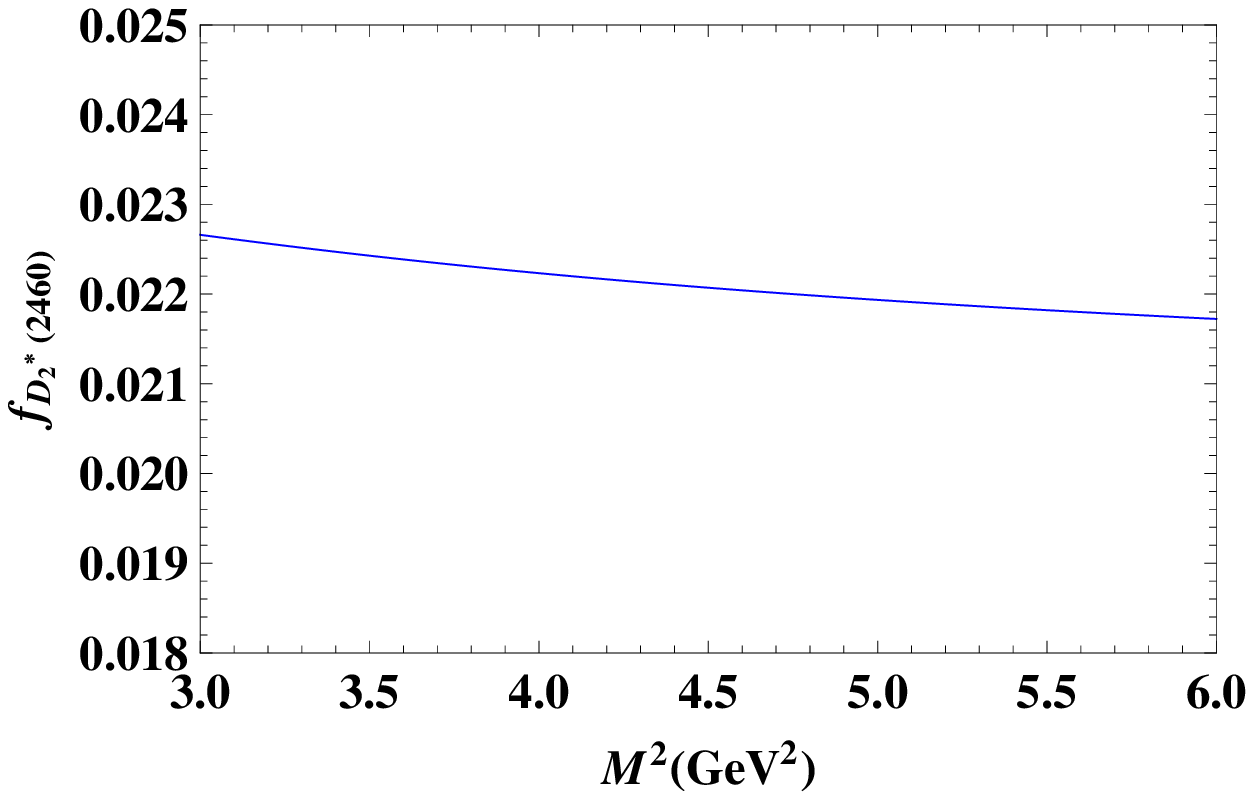}
\end{center}
\caption{Dependencies of the mass and meson-current coupling constant of the $D_2^*(2460)$ tensor meson on Borel mass parameter in its working region.}
\label{Diagrams1}
\end{figure}
 Our numerical results on the
mass and meson-current coupling constant for $D_2^*(2460)$ tensor meson as well as
the experimental data on the mass \cite{K.Nakamura} are given in
Table \ref{tab1}.
\begin{table}[h] \centering
 \begin{tabular}{|c||c|c|c|c|} \hline &
 Present Work&Experiment \cite{K.Nakamura}
\\\cline{1-3}\hline\hline
$m_{D_2^*(2460)}$& $(2.53\pm0.45)~GeV $ &$(2.4626\pm0.0007)~GeV$\\
\cline{1-3} $f_{D_2^*(2460)}$& $0.0228\pm0.0068$ &-\\
\cline{1-3}\hline\hline
\end{tabular}
\vspace{0.8cm} \caption{Values for the mass and meson-current coupling constant of
the $D_2^*(2460)$ tensor meson.} \label{tab1}
\end{table}
The errors quoted in our predictions are due to the variations of both
auxiliary parameters and uncertainties in input parameters. From Table \ref{tab1}, we see a good consistency
between our prediction and  the experimental data
on the mass of the $D_2^*(2460)$ tensor meson. Any measurement on the meson-current coupling constant and comparison of the result with our prediction can give more information about the nature of the
orbitally excited $D_2^*(2460)$ tensor meson.

\section{Acknowledgment}
We thank Y. Sara\c{c} for  useful discussions. This work has been
supported in part by the scientific and technological research council of
Turkey (TUBITAK) under research project No. 110T284.

\end{document}